\documentclass[twocolumn,amsmath,amssymb,aps,secnumarabic,%
    nofootinbib,superscriptaddress]{revtex4}
\usepackage{times,mathptmx,amsthm}
\newtheorem{theorem}{Theorem}
\newtheorem{lemma}[theorem]{Lemma}

\theoremstyle{remark}
\newtheorem*{remark}{Remark}

\renewcommand{\tensor}{\otimes}
\newcommand{\braket}[1]{{\langle #1 \rangle}}
\renewcommand{\hat}{\widehat}
\renewcommand{\tilde}{\widetilde}
\newcommand{\sa}{\mathrm{sa}}
\DeclareMathOperator{\Tr}{Tr}
\newcommand{\eps}{\epsilon}

\newcommand{\ie}{\textit{i.e.}}
\newcommand{\Ie}{\textit{I.e.}}

\newcommand{\C}{\mathbb{C}} \newcommand{\R}{\mathbb{R}}
\newcommand{\Z}{\mathbb{Z}}

\newcommand{\cA}{\mathcal{A}} \newcommand{\cB}{\mathcal{B}}
\newcommand{\cL}{\mathcal{L}} \newcommand{\cM}{\mathcal{M}}

\newcommand{\tA}{\tilde{A}} \newcommand{\tB}{\tilde{B}}
\newcommand{\tC}{\tilde{C}}

\newcommand{\vA}{{\vec{A}}} \newcommand{\vB}{{\vec{B}}}
\newcommand{\vC}{{\vec{C}}} \newcommand{\vW}{{\vec{W}}}
\newcommand{\vX}{{\vec{X}}} \newcommand{\vY}{{\vec{Y}}}
\newcommand{\vZ}{{\vec{Z}}}

\newcommand{\valpha}{{\vec{\alpha}}} \newcommand{\vbeta}{{\vec{\beta}}}

\newcommand{\vt}{{\vec{t}}} \newcommand{\vx}{{\vec{x}}}
\newcommand{\vy}{{\vec{y}}}

\newcommand{\tvA}{\tilde{\vA}} \newcommand{\tvB}{\tilde{\vB}}
\newcommand{\tvC}{\tilde{\vC}}

\newcommand{\hcL}{\hat{\mathcal{L}}}
\newcommand{\hf}{\hat{f}} \newcommand{\hW}{\hat{W}}

\newcommand{\eatline}{\vspace{-\baselineskip}}
\newcommand{\eathalfline}{\vspace{-0.5\baselineskip}}

\newcommand{\thm}[1]{Theorem~\ref{#1}}

\newcommand{\lem}[1]{Lemma~\ref{#1}}
\newcommand{\eq}[2]{\begin{equation}\label{#1}#2\end{equation}}

\begin{document}
\title{A tracial quantum central limit theorem}

\author{Greg Kuperberg}
\email{greg@math.ucdavis.edu}
\thanks{Supported by NSF grant DMS \#0072342}
\affiliation{UC Davis}

\begin{abstract}
We prove a central limit theorem for non-commutative random variables in a von
Neumann algebra with a tracial state:  Any non-commutative polynomial of
averages of i.i.d. samples converges to a classical limit.  The proof is based
on a central limit theorem for ordered joint distributions together with
a commutator estimate related to the Baker-Campbell-Hausdorff expansion. The
result can be considered a generalization of Johansson's theorem on the
limiting distribution of the shape of a random word in a fixed alphabet as its
length goes to infinity.
\end{abstract}
\maketitle

\section{Introduction}

One of the most important results in mathematics for science is the central
limit theorem.  But there is only an incomplete theory of central limits in the
setting of quantum probability theory, even though the quantum rules of
probability describe physical reality just as the classical rules do.  In this
paper we will prove a quantum central limit theorem for tracial states.  Our
result is not as sharp or as general as what one may conjecture, but it is
sharper than previous central limits theorems in the same setting
\cite{GVV:central,CH:mechanical,Quaegebeur:ccr,GW:algebraic}.

The difference between quantum and classical central limits only appears in the
multivariate case.  In quantum probability theory, a real-valued random
variable is expressed by a self-adjoint operator in a von Neumann algebra (see
below). Quantum behavior only arises with non-commuting operators.  In
particular independent variables do commute, so the proof of the classical
central limit theorem applies to independent, identically distributed (i.i.d.)
real-valued quantum random variables.

The classical theorem does not apply to i.i.d. samples of a vector of random
variables with non-commuting components.  For example, let $A$ and $B$ be two
non-commuting measurements in a quantum probability space $\cM$, and let $\tA$
and $\tB$ be the corresponding averaged measurements in the larger space
$\cM^{\tensor N}$ consisting of $N$ independent copies of $\cM$.  We would like
to say that the joint distribution of $\tA$ and $\tB$ converges to a Gaussian
distribution.  But since $\tA$ and $\tB$ are not simultaneously measurable, it
is not clear what this means.  One physically observable interpretation is that
any suitable self-adjoint expression, such as the anti-commutator $\tA\tB +
\tB\tA$, should approach a classical distribution.

Our result follows this interpretation.  To state it precisely we introduce
some notation.  Let $(\cM,\rho)$ a quantum probability space consisting of a
von Neumann algebra $\cM$ and a normal state $\rho$. Then $(\cM^{\tensor
N},\rho^{\tensor N})$ denotes $N$ independent copies of $(\cM,\rho)$.  If $A
\in \cM$, then
$$A^{(k)} = I^{\tensor k-1} \tensor A \tensor I^{\tensor N-k}$$
denotes the $k$th copy of $A$ in $\cM^{\tensor N}$, while
$$\tA = \frac{A^{(1)} + A^{(2)} + \cdots + A^{(N)}}{\sqrt{N}}$$
is the rescaled sum of $N$ independent samples of $A$. Finally if $A$ is a
quantum random variable, $\cL[A]$ denotes its distribution, or spectral measure.

\begin{theorem} Let $(\cM,\rho)$ be a quantum probability space with a tracial
state $\rho$, and let $A_1,A_2,\ldots,A_a$ be self-adjoint elements with mean
0. Let $p \in \C\braket{A_1,A_2,\ldots,A_a}$ be a self-adjoint non-commutative
polynomial in $k$ variables. Then
$$\lim_{N \to \infty} \cL[p(\tA_1,\tA_2,\ldots,\tA_a)] =
    \cL[p(X_1,X_2,\ldots,X_a)],$$
where $X_1,X_2,\ldots,X_a$ are classical Gaussian random variables with
covariance matrix
$$E[X_j X_k] = \rho(A_j A_k).$$
\label{th:main}
\eatline\end{theorem}

We briefly review the terminology and notation in \thm{th:main} and the
discussion above \cite{KR:vol1,KR:vol2}. A \emph{von Neumann algebra} is an
algebra of bounded operators on some Hilbert space which is both $*$-closed and
closed in the strong (equivalently weak) operator topology.  A \emph{state}
$\rho$ is a bounded, $*$-invariant functional on $\cM$ which is non-negative on
positive elements of $\cM$; it is called \emph{normal} if it is continuous with
respect to the strong (equivalently weak) operator topology.  A state $\rho$ is
\emph{tracial} if
$$\rho(AB) = \rho(BA)$$
for all $A,B \in M$.  If $A \in \cM$ is self-adjoint, then any normal state
$\rho$ on $\cM$ induces a measure on the spectrum of $A$, the (scalar-valued)
\emph{spectral measure} on $A$.  Thus we can interpret $A$ as a real-valued
random variable, this is the Copenhagen or Born interpretation in physics.  If
$\cM$ is commutative, then it is analogous to a $\sigma$-algebra, while $\rho$
is analogous to a measure on the algebra.

In physics terminology, $\cM$ is a suitably complete calculus of operators on a
quantum system.  The state $\rho$ then corresponds to a density operator or an
ensemble, although the mathematical convention is to express it as a linear
functional on operators rather than as an operator. \Ie, $\rho(A)$ denotes the
expectation or mean of $A$ with respect to $\rho$, often written $\Tr(\rho A)$
in physics.

The ring $\C\braket{A_1,\ldots,A_a}$ of non-commutative polynomials admits a
unique $*$-involution which fixes each $A_n$ and which, as usual, is
anti-linear and an algebra anti-automorphism. A polynomial $p$ is self-adjoint
if $p^* = p$.  For example, the commutator $[A_1,A_2]$ is anti-self-adjoint,
but $i[A_1,A_2]$ is self-adjoint.

Readers who aren't interested in general von Neumann algebras can consider
the special case $\cM = \cM_k$, the $k \times k$ matrices, and
$$\rho(A) = \frac{\Tr(A)}{k}.$$
In this case each $A_n$ is a $k \times k$ Hermitian matrix.  We interpret a
Hermitian matrix $A$ as a random variable by the formula
$$P[A = \lambda] = \frac{m}{k}$$
if $\lambda$ is an eigenvalue of $A$ with multiplicity $m$.  This
special case captures the difficulties of the general case.
We refer readers to Nielsen and Chuang \cite{NC:book} or Sakurai
\cite{Sakurai:modern} for introductions to quantum probability theory.

With these definitions we can discuss the limitations of \thm{th:main}.
Possibly the most serious one is that $\rho$ must be tracial.  We previously
conjectured \thm{th:main} without assuming that $\rho$ is tracial, only that
the covariance matrix of $A_1,A_2,\ldots,A_a$ is symmetric
\cite{Kuperberg:words}.  Goderis, Verbeure, and Vets found a quantum
central limit for any state $\rho$, namely a quasi-free state (or in
physics terminology, a product of thermal states of harmonic oscillators)
\cite{GVV:central}. Thus we conjecture that in \thm{th:main} we can let
$\rho$ be any normal state if we replace ``classical Gaussian random
variables'' by ``quasi-free variables''.

Of course \thm{th:main} applies to the case where $\rho$ does not symmetrize
all of $\cM$, only the subalgebra generated by $A_1,\ldots,A_a$, since we can
replace $\cM$ by this subalgebra. In physics language it suffices for
$A_1,\ldots,A_a$ to commute with the density matrix.  This occurs, for example,
if $\rho$ is a thermal state (also called a Gibbs state or the Boltzmann
distribution) of a Hamiltonian $H$ and each $A_n$ is a conserved quantity.  It
also occurs in the infinite-temperature limit of any system, because in this
limit the density matrix approaches the identity.

Another limitation of \thm{th:main} is the requirement that each $A_n$ is
bounded.  By contrast the classical central limit theorem requires only that
the covariance matrix is finite.  Thus a more satisfying version would allow
each $A_n$ to be an unbounded operator affiliated with $\cM$, although with
enough restrictions that $p(\tA_1,\ldots,\tA_a)$ is still well-defined.
Finally the polynomial ring $\C\braket{A_1,\ldots,A_a}$ could be replaced by
some von Neumann algebra to which $\C\braket{A_1,\ldots,A_a}_\sa$ is
affiliated.  This algebra would model measurable non-polynomial expressions in
the variables $A_1,\ldots,A_a$; it would be a non-commutative analogue of the
algebra $L^\infty(\R^a)$ of bounded, measurable functions on $\R^a$.

Despite limitations, \thm{th:main} is useful.  For instance, we previously
showed that it implies Johansson's theorem on the limiting distribution of
the shape of a random word in $k$ letters as the length $N$ goes to infinity
\cite{Kuperberg:words,Biane:dual,Johansson:plancherel}. Indeed \thm{th:main}
can be taken as a generalization of Johansson's theorem.

\subsection{Some previous results}
\label{s:prev}

Many results in the literature are called or could be called quantum central
limit theorems.  Here we discuss a few of that apply to a discrete set of
independent, non-commuting random variables, as \thm{th:main} does.

Cushen and Hudson proved the quantum central limit theorem for a pair of
conjugate variables $P$ and $Q$, \ie, such that $[P,Q] = iI$
\cite{CH:mechanical}. Quaegebeur later generalized this result to arbitrary
CCR algebras \cite{Quaegebeur:ccr}.

Giri and von Waldenfels proved that the general non-commutative moment
$\rho(A_1A_2\ldots A_a)$ converges to a Gaussian or quasi-free
value \cite{GW:algebraic}.  This can be considered an ``algebraic''
quantum central limit theorem because it applies to any $*$-algebra.

Goderis, Verbeure, and Vets established the convergence of $\rho^{\tensor
N}(e^{i\tA})$ for all $A \in \cM$ to the characteristic functions of a
quasi-free state of a universal CCR-algebra over $\cM$ \cite{GVV:central}.

If we compare these three results  and \thm{th:main} to each other,
none is eclipsed by the others.  As implied above, we conjecture that there is
a mutual generalization of all four results in the von Neumann algebra setting.

Finally Voiculescu and many followers have developed a non-commutative
probability theory in which conventional statistical independence is
replaced by free independence in the sense of non-commutative free products
\cite{Voiculescu:lectures}.  This is a very interesting theory which may yet
be useful in physics, but the results in this paper are not part of it.

\acknowledgments

We especially thank Bruno Nachtergaele for his continued attention to this
work.  We also thank Philippe Biane, Janko Gravner, Tom Michoel, Marc
Rieffel, and Dan Voiculescu for useful comments.

\section{Outline of the proof}

We use $\cM_\sa$ to denote the space of self-adjoint elements of $\cM$.  We
abbreviate $\vA = (A_1,A_2,\ldots,A_a)$. We also abbreviate
$$\vA^\valpha = (A_1^{\alpha_1},A_2^{\alpha_2},\ldots,A_a^{\alpha_a}),$$
and we use $\vx \cdot \vy$ to denote the standard inner product on $\R^a$.

Our proof of \thm{th:main} is based on a multivariate generalization of the
spectral measure $\cL[A]$ that we call the ordered joint distribution
$$\cL[A_1,A_2,\ldots,A_a] = \cL[\vA].$$
It can be defined by its Fourier transform
$$\hcL[\vA](\vt) = \rho(e^{it_1 A_1}e^{it_2 A_2}\cdots
     e^{t_k A_a^{\alpha_a}}).$$
The ordered joint distribution is not directly observable, among other
reasons because it allows negative and even non-real ``probabilities''.

As a first step, an essentially classical central limit theorem holds for
ordered joint distributions.  Consequently for any positive exponents
$\alpha_1,\alpha_2,\ldots,\alpha_a$, the distribution
$$\cL[\tA_1^{\alpha_1},\tA_2^{\alpha_2},\ldots,\tA_a^{\alpha_a}] =
    \cL[\tvA^\valpha]$$
also approaches a classical limit.  When $\rho$ is
tracial, the difference between an ordered joint characteristic function,
$$\hcL[\vA](\vt) = \rho(e^{it_1 \tA_1^{\alpha_1}}e^{it_2 \tA_2^{\alpha_2}}
    \cdots e^{it_k\tA_a^{\alpha_a}}),$$
and the characteristic function of a linear combination of
powers of $\tA_1,\tA_2,\ldots,\tA_a,$
$$\rho(e^{i(t_1 \tA_1^{\alpha_1}+t_2\tA_2^{\alpha_2}+\cdots+
    t_k \tA_a^{\alpha_a})}) = \rho(e^{i \vt \cdot \tvA^\valpha}),$$
is bounded by a decaying commutator estimate.  Thus the spectral
characteristic functions converge pointwise to a classical limit when
$p(\vA)$ is a linear combination of powers.  This implies many special cases
of \thm{th:main}, indeed every case up to a change of variables.

\section{The proof}

The proof is more natural in the $C^*$-algebra setting than the von Neumann
algebra setting.  Recall that a $C^*$-algebra is special kind of complex Banach
algebra and that every von Neumann algebra is a $C^*$-algebra \cite{KR:vol1}.
$C^*$-algebras have states, but they do not have normal states in the absence
of weak and strong operator topologies.  By the GNS construction, given any
state $\rho$ on a $C^*$-algebra $\cA$, there exists a von Neumann algebra $\cM
\supseteq \cA$ and an extension of $\rho$ which is normal on $\cM$. Henceforth
we replace the von Neumann algebra $\cM$ by a $C^*$-algebra $\cA$ and drop the
inessential condition that $\rho$ is normal. (This is analogous to considering
probability distributions on topological spaces instead of measure spaces.)

We call a norm $||\cdot||_S$ on a $C^*$-algebra $\cA$ \emph{spectral}
if it is $*$-invariant and if
$$||UA||_S = ||A||_S$$
for any unitary $U$.  We consider the GNS norm
$$||A||_\rho = \sqrt{\rho(A^*A)}$$
on $\cA$, which is spectral when $\rho$ is tracial.  (If $\rho$ is not
faithful, it is only a semi-norm.)  By the Russo-Dye theorem \cite{KR:vol2},
$$||AB||_S \le \min\{||A||_S\;||B||,\;||A||\;||B||_S\}$$
in any spectral norm $||\cdot||_S$.  In particular
$$||A||_S \le ||1||_S\;||A||,$$
so the topology induced by $||A||_S$ is at least as coarse as the norm
topology.

If $\cA$ happens to be a von Neumann algebra and $A \in \cA_\sa$ has a discrete
spectrum, then its spectral measure $\cL[A]$ is given by the rule
$$P[A = \lambda] = \rho(A_\lambda),$$
where $\lambda$ is an eigenvalue of $A$ and $A_\lambda$ denotes projection onto
its $\lambda$-eigenspace.  If $A_1,A_2,\ldots,A_a \in \cA_\sa$ all have
discrete spectra, the \emph{ordered joint distribution} $\cL[\vA]$ is likewise
given by
$$P[\vA = \vec{\lambda}] = \rho((A_1)_{\lambda_1}(A_2)_{\lambda_2}
    \cdots (A_a)_{\lambda_a}).$$
This is generally a complex-valued measure rather than a non-negative measure.
Without the discrete spectrum assumption, $\cL[\vA]$ might not strictly be a
measure at all.  Recall that a probability measure on $\R^a$ can be regarded as
a bounded functional on the space $C_0(\R^a)$ of continuous functions on $\R^a$
that vanish at infinity.  In this paper we define $\cL[\vA]$ as a function on
the space of products
$$f(\vx) = f_1(x_1) f_2(x_2) \cdots f_a(x_a)$$
of decaying, continuous, univariate functions by the rule
$$\cL[\vA](f) = \rho(f_1(A_1)f_2(A_2) \cdots f_a(A_a)).$$
Since $\rho$ is bounded, $\cL[\vA]$ is a bounded functional on the projective
Banach tensor product $C_0(\R)^{\tensor a}$ (\ie, the minimal completion of the
algebra tensor product, or  the completion with respect to the greatest cross
norm).  In order to be a genuine measure, $\cL[\vA]$ would have to extend from
$C_0(\R)^{\tensor a}$ to the much larger Banach space $C_0(\R^a)$. This is not
always possible, although $\cL[\vA]$ might naturally extend to some
intermediate Banach space. The distribution $\cL[\vA]$ is also sensitive to
permutations of the variables.  But it otherwise reasonably generalizes the
usual joint distribution of classical random variables.

\begin{remark}
For conjugate variables $P$ and $Q$, the ordered joint distribution is related
to the Wigner distribution.  The Fourier transform of the Wigner distribution
is given by
$$\hW(q,p) = \rho(e^{iPq/2} e^{iQp} e^{iPq/2}).$$
We can then compare
\begin{align*}
\hcL[P,Q](q,p) &= \rho(e^{iPq} e^{iQp}) = e^{-ipq/2} \hW(q,p) \\
\hcL[Q,P](p,q) &= \rho(e^{iQp} e^{iPq}) = e^{ipq/2}  \hW(q,p).
\end{align*}
Thus the Fourier transform of the Wigner distribution is a phase-corrected form
of the Fourier transform of the ordered joint distribution.
\end{remark}

\begin{theorem} If $\cA$ is a $C^*$-algebra with
a tracial state $\rho$ and $A_1,A_2,\ldots,A_a \in \cA_\sa$ with mean zero,
then their ordered joint distribution $\cL[\vA]$ obeys the central limit
theorem:
$$\lim_{N \to \infty} \cL[\tvA] = \cL[\vX]$$
weakly as functionals on $C_0(\R)^{\tensor a}$, and $\vX$ is Gaussian with
covariance matrix
$$M_{j,k} = E[X_j X_k] = \rho(A_j A_k).$$
\eatline \label{th:ojcentral} \end{theorem}

\begin{proof} We follow standard proofs of the central limit theorem by the
method of characteristic functions \cite{BR:normal}, but the argument must be
applied carefully because $\cL[\vA]$ is a complex measure.  We will argue
convergence of $\cL[\vA]$ on four classes of functions in turn:
\begin{description}
\item[1.] products of sinusoids, \eathalfline
\item[2.] products of smooth functions with bounded support, \eathalfline
\item[3.] products of bounded, smooth functions, and \eathalfline
\item[4.] products of bounded, continuous functions.
\end{description}

Observe that if $B_1,B_2,\ldots,B_a$ are self-adjoint elements
in another $C^*$-algebra $\cB$ with a state $\tau$ and
$$A_n^{(1)} = A_n \tensor I \in \cA \tensor \cB \qquad
B_n^{(2)} = I \tensor B_n \in \cA \tensor \cB,$$
then
$$\cL[\vA^{(1)} + \vB^{(2)}] = \cL[\vA] * \cL[\vB],$$
where the operation $*$ denotes convolution.  Thus
\eq{e:conv}{\cL[\tvA] = T_{1/\sqrt{N}}(\cL[\vA] * \cL[\vA] * \cdots *
    \cL[\vA]),}
where $T_x$ denotes the operator on functionals induced by rescaling space by a
factor of $x$.  Equation~\eqref{e:conv} implies convergence of characteristic
functions:
$$\hcL[\tvA](\vt) = \rho^{\tensor N}(e^{i\vt \cdot \tvA}) \to
    e^{-\vt \cdot M \vt/2}$$
locally uniformly in $\vt$.

Suppose that
$$f(\vt) = f_1(t_1) f_2(t_2) \cdots f_a(t_a)$$
is a product of smooth univariate functions with bounded support.
Its Fourier transform $\hf$ decays super-polynomially.  Consider
the joint expectation
$$\cL[\tvA](f) = (2\pi)^{-a/2}\int_{\R^a} \hcL[\tvA](\vt)
    \overline{\hf(\vt)} d\vt$$
as an integral in Fourier space.  For any $R>0$, the integral converges
inside the box $[-R,R]^a$ because there the integrand converges uniformly in
$\vt$.  On the other hand, the integrand vanishes outside of the box
uniformly in $N$ as $R \to \infty$ because $\hf$ decays and for all $\vt$,
$$|\hcL[\tvA](\vt)| = |\rho^{\tensor N}(e^{i\vt \cdot \tvA})| \le 1.$$
(Both $\rho^{\tensor N}$ and its argument have norm 1.)
Thus for these $f$,
$$\cL[\tvA](f) \to \cL[\vX](f).$$

Now suppose that $f_1,f_2,\ldots,f_a$ have bounded support but are
merely continuous.  By the Weierstrass approximation theorem,
for each $\eps > 0$, we can let
$$f_n(t) = f_n^{(1)}(t) + f_n^{(2)}(t),$$
where $f_n^{(1)}(t)$ is smooth and
$$||f_n^{(2)}|| \le \eps.$$
Then
$$\cL[\tvA](f) = \sum_{\vec{\sigma} \in \{1,2\}^a} \hspace{-.7em}
    \rho^{\tensor N}(f^{(\sigma_1)}_1(\tA_1)f^{(\sigma_2)}_2(\tA_2)
    \cdots f^{(\sigma_a)}_a(\tA_a)).$$
Each term of the sum other than the first one is bounded by $\eps$,
while the first term converges to $\cL[\vX](f)$ if we take $\eps \to 0$.
Thus for continuous $f$ with bounded support,
$$\cL[\tvA](f) \to \cL[\vX](f).$$

Finally suppose that $f_1,f_2,\ldots,f_a$ are bounded but do not
have bounded support.  Without loss of generality we suppose that
$$||f_n|| \le 1$$
for each $n$. For each $R > 0$, we can choose a continuous partition
$$f_n(t) = f_n^{(1)}(t) + f_n^{(2)}(t),$$
where
$$f_n^{(1)}(t) = \min\{\max\{0,|t|-R\},1\}f_n(t).$$
Thus
$$f_n(t) = \begin{cases} f_n^{(1)}(t) & |t| \le R \\
    f_n^{(2)}(t) & |t| \ge R+1 \end{cases}. $$
By the univariate central limit theorem,
$$\lim_{R \to \infty}||f_n^{(2)}(\tA_n)||_{\rho^{\tensor N}} = 0$$
uniformly in $N$.
Let $\eps>0$ and choose $R$ such that for each $n$,
$$||f_n^{(2)}(\tA_n)||_{\rho^{\tensor N}} < \eps.$$
Since for each $n$,
$$||f_1^{(2)}(\tA_n)|| \le 1,$$
and since $||\cdot||_{\rho^{\tensor N}}$ is spectral, each
term other than the first in the expansion
$$\cL[\tvA](f) = \sum_{\vec{\sigma} \in \{1,2\}^a} \hspace{-.7em}
    \rho^{\tensor N}(f^{(\sigma_1)}_1(\tA_1)f^{(\sigma_2)}_2(\tA_2)
    \cdots f^{(\sigma_a)}_a(\tA_a))$$
is bounded by $\eps$. (The $\rho^{\tensor N}$-norm of the argument in each term
is bounded by $\eps$.)  As before, the first term converges to $\cL[\vX](f)$ if
we take $\eps \to 0$. Thus for all $f$ described in the theorem,
$$\cL[\tvA](f) \to \cL[\vX](f).$$
\end{proof}

\begin{lemma} If $A,B$ are elements of a $C^*$-algebra $\cA$, then
$$e^A e^B - e^{A+B} = \int_0^1 \int_0^{1-t}
    e^{t(A+B)}e^{(1-t-s)A} [A,B] e^{sA} e^{(1-t)B} ds\;dt.$$
\eatline \label{l:cint} \end{lemma}

\begin{proof} Let
$$X = e^{A/n} \qquad Y = e^{B/n}.$$
By elementary calculation,
\eq{e:csum}{X^nY^n - (XY)^n = \sum_{t=1}^{n-1} \sum_{s=0}^{n-t-1}
    (XY)^{t-1} X^{n-t-s} [X,Y] X^s Y^{n-t}.}
Since
\begin{align*}
\lim_{n \to \infty} X^{tn} &= e^{tA}
    & \lim_{n \to \infty} Y^{tn} &= e^{tB} \\
\lim_{n \to \infty} (XY)^{tn} &= e^{t(A+B)}
    & \lim_{n \to \infty} n^2[X,Y] &= [A,B],
\end{align*}
equation~\eqref{e:csum} converges to the statement of the lemma
as $n$ goes to infinity.
\end{proof}

\begin{lemma} If $A_1,A_2,\ldots,A_a$ are self-adjoint elements of a
$C^*$-algebra $\cA$ and $||\cdot||_S$ is a spectral semi-norm, then
$$||e^{iA_1}e^{iA_2}\cdots e^{iA_a} - e^{i(A_1 + A_2 + \cdots + A_a)}||_S
    \le \sum_{1 \le j<k \le a} \hspace{-1em}\frac{||[A_j,A_k]||_S}{2}.$$
\eatline \label{l:spec} \end{lemma}

\begin{proof} If $a = 2$, then \lem{l:cint} implies that
$$|| e^{iA_1}e^{iA_2} - e^{i(A_1 + A_2)}||_S \le \frac{||[A_1,A_2]||_S}2$$
by taking the norm inside the integral.  Note that the integral is defined
convergence in the algebra norm, but it is equally valid to exchange it
with any other form of convergence which is at least as weak, as convergence
in $||\cdot||_S$ is necessarily.

The general case follows by induction.
\end{proof}

Let $\C\{\vA\}$ denote the Lie algebra freely generated by the symbols
$A_1,A_2,\ldots,A_a$.  Recall that its universal enveloping algebra is the ring
of non-commutative polynomials in the same variables:
$$U(\C\{\vA\}) = \C\braket{\vA}.$$

\begin{lemma} If $X$ and $Y$ commute, $\alpha,\beta \in \Z_{\ge 0}$, and $q >
1$, then
$$X^\alpha Y^\beta = \sum_{n=0}^{\alpha+\beta} t_n (X+q^n Y)^{\alpha+\beta}$$
for some $t_0,t_1,\ldots,t_{\alpha+\beta} \in \R$.
\label{l:powsum}
\end{lemma}
\begin{proof}
The coefficients $t_0,t_1,\ldots,t_{\alpha+\beta}$ must satisfy
the linear system
$$\sum_{n=0}^{\alpha+\beta} q^{nk} =
    \begin{cases} \frac1{\binom{\alpha+\beta}{\alpha}} & k = \alpha \\
    0 & k \ne \alpha \end{cases}$$
for $0 \le k \le \alpha+\beta$.  The matrix of this system is a Vandermonde
matrix, invertible when $q>1$, so the system has a solution.
\end{proof}

\begin{lemma} Any $p \in \C\braket{\vA}$ can be expressed as a linear
combination of power sums
$$p(\vA) = t_1 B_1^{\beta_1} + t_2 B_2^{\beta_1} + \cdots
    + t_b B_b^{\beta_b} = \vt \cdot \vB^\vbeta$$
with each $B_n \in \C\{A_1,A_2,\ldots,A_a\}$ and each $\beta_n \in \Z_{\ge 0}$.
If $p$ is self-adjoint, then we can take each $t_n \in \R$ and each $B_n$ to be
self-adjoint.
\label{l:lie} \end{lemma}

\begin{proof} The Poincar\'e-Birkhoff-Witt theorem provides a vector
space isomorphism
$$\Phi:S(L) \to U(L)$$
from the symmetric algebra to the universal enveloping algebra of any Lie
algebra $L$ \cite{Humphreys:gtm}.  The map $\Phi$ is given by symmetrization:
$$\Phi(X_1X_2\cdots X_n) = \frac1{n!}\sum_{\pi \in S_n}
    X_{\pi(1)}X_{\pi(2)}\cdots X_{\pi(n)}$$
for any $X_1,X_2,\ldots,X_n \in L$.  If $L$ has an anti-involution $*$, then it
extends to both $S(L)$ and $U(L)$ and $\Phi$ intertwines it. The map $\Phi$
also preserves powers of elements of $L$:
$$\Phi(X^n) = X^n.$$
It therefore also preserves linear combinations of powers.

We claim that if $L$ is a vector space, then every element of the symmetric algebra
$S(L)$ is a linear combination of powers.  \lem{l:powsum} establishes the
special case that the product of two powers is a linear combination of powers.
The set $P$ of linear combinations of powers is therefore closed under
multiplication; it is a (complex) subalgebra of $S(L)$.  Since $L \subset P$
trivially and $L$ generates $S(L)$, we conclude that $P = S(L)$.  If $L$ has an
anti-linear $*$-involution, then the same argument applies to $S(L)_\sa$, which
is generated as a real algebra by the real vector space $L_\sa$.

In conclusion, for any Lie algebra $L$, powers of elements of $L$ span $U(L)$
as a complex vector space.  If $L$ has an anti-involution $*$, then powers of
elements of $L_\sa$ span $U(L)_\sa$ as a real vector space.  The lemma is the
special case
$$L = \C\{A_1,A_2,\ldots,A_a\}.$$
\end{proof}

The proof of \thm{th:main} is simpler if the vector of Lie elements $\vB$
provided by \lem{l:lie} is linear in $\vA$.  (This includes many interesting
choices for $p$, for instance the anti-commutator $A_1 A_2 + A_2 A_1$.)  Indeed
in this case
$$p(\tvA) = \vt \cdot \tvB^\vbeta$$
is a polynomial in $\tvB$, so we could replace $\vA$ with $\vB$ in the
statement of the theorem.

\begin{lemma} If $A,B \in \cA_\sa$ have mean 0 and $\alpha,\beta \in
\Z_{\ge 0}$, then
$$||[\tA^\alpha,\tB^\beta]||_{\rho^{\tensor N}} = O(N^{-1/2}) ||A||\;||B||.$$
\eatline \label{l:cnorm} \end{lemma}

Before proving \lem{l:cnorm} in full generality, we motivate it with a simple
proof when $\alpha = \beta = 1$.  In this case
$$[\tA^\alpha,\tB^\beta] = \frac{1}{\sqrt{N}} \tilde{[A,B]}.$$
Since $[A,B]$ has mean 0 (because $\rho$ is tracial), the typical eigenvalue of
$\tilde{[A,B]}$ is $O(1)$.  Thus the norm of the right side is $O(N^{-1/2})$.

\begin{proof} In brief, since
$$\rho(A) = \rho(B) = \rho([A,B]) = 0,$$
and since $\rho$ is tracial, the multilinear expansion of
$$\rho^{\tensor N}([\tA^\alpha,\tB^\beta]^2) =
    \pm ||[\tA^\alpha,\tB^\beta]||^2_{\rho^{\tensor N}}$$
has $O(N^{\alpha+\beta-1})$ non-cancelling terms. Since the expansion also has
a factor of $N^{\alpha+\beta}$ in the denominator, the square of the norm is
$O(N^{-1})$.

In detail, the expansion is
\begin{multline}
N^{\alpha+\beta}\rho^{\tensor N}([\tA^\alpha,\tB^\beta]^2) = \\
\sum_{\vX,\vY,\vZ,\vW} \biggl(
\prod_n \rho(X_n Y_n Z_n W_n)
- \prod_n \rho(Y_n X_n Z_n W_n) \\
- \prod_n \rho(X_n Y_n W_n Z_n)
+ \prod_n \rho(Y_n X_n W_n Z_n)\biggr),
\label{e:expand}
\end{multline}
where each vector $\vX$ and $\vZ$ consists of $\alpha$ copies of $A$ and
$N-\alpha$ copies of $I$, and each vector $\vY$ and $\vW$ consists of $\beta$
copies of $B$ and $N-\beta$ copies of $I$.  Momentarily fix $\vX,\vY,\vZ,\vW$
and consider the corresponding four terms in equation~\eqref{e:expand}. If for
some $n$, exactly one of $X_n, Y_n, Z_n, W_n$ is $A$ or $B$, then all four
terms vanish.  Likewise if for all $n$, not all four of $X_n, Y_n, Z_n, W_n$
are $A$ or $B$, then the four terms cancel.  Thus in a non-cancelling choice of
the vectors $\vX,\vY,\vZ,\vW$, the four components $X_n,Y_n,Z_n,W_n$ are
non-trivial for at most $\alpha+\beta-1$ values of $n$.  There are only
$O(N^{\alpha+\beta-1})$ such terms, each bounded by $||A||^2 ||B||^2$.
\end{proof}

Applying \thm{th:ojcentral} to $\vB$, we learn not only that
$$\lim_{N \to \infty} \cL[\tvB] = \cL[\vY]$$
with $\vY = (Y_1,Y_2,\ldots,Y_b)$ Gaussian, but also that
$$\lim_{N \to \infty} \cL[\tB^\vbeta] = \cL[\vY^\vbeta].$$
Thus for every $\vt$,
\eq{e:ojpow}{\hcL[\tB^\vbeta](\vt) =
    \rho^{\tensor N}(e^{t_1 \tB_1^{\beta_1}} e^{it_2 \tB_2^{\beta_2}} \cdots
    e^{it_b \tB_b^{\beta_b}}) \to \hcL[\vY^\vbeta](\vt)}
approaches a classical limit as $N \to \infty$.  Combining
\lem{l:spec} with \lem{l:cnorm},
\eq{e:reorder}{||e^{it_1 \tB_1^{\beta_1}} e^{it_2 \tB_2^{\beta_2}} \cdots
    e^{it_b \tB_b^{\beta_b}} - e^{i\vt \cdot \tvB^\vbeta}||_{\rho^{\tensor N}}
    = O(N^{-1/2}).}
Finally combining equations~\eqref{e:ojpow} and \eqref{e:reorder} with
$$||\rho^{\tensor N}||_{\rho^{\tensor N}} = 1,$$
we obtain
$$\rho^{\tensor N}(e^{i\vt \cdot \tvB^\vbeta}) \to \hcL[\vY^\vbeta](\vt).
$$
Replacing $\vt$ by $z\vt$, we obtain
$$\hcL[p(\tvA)](z) = \rho^{\tensor N}(e^{iz\vt \cdot \tvB^\vbeta})\to
\hcL[\vt \cdot \vY^\vbeta](z).$$
Since $p(\vX) = \vt \cdot \vY^\vbeta$ and pointwise convergence of
characteristic functions implies weak convergence of measures, we have
established \thm{th:main} when $\vB$ is linear.

The idea behind the general case is that the non-linear terms in each $B_n$
decay as $N \to \infty$. We expand each $B_n$ as a sum of homogeneous terms:
$$B_n = B_{n,1} + B_{n,2} + \ldots + B_{n,d_n},$$
where $B_{n,d}$ is the degree $d$ term of $B_n$. Then
$$p(\tA_1,\tA_2,\ldots,\tA_a) = t_1 \tC_1^{\beta_1} + t_2 \tC_2^{\beta_1}
    + \cdots + t_b \tC_b^{\beta_b},$$
where
$$
C_n = B_{n,1} + N^{-1/2} B_{n,2} + \ldots + N^{(1-d_n)/2} B_{n,d_n}.
$$
Furthermore
$$\rho(B_{n,1}) = \rho(B_{n,2}) = 0,$$
in the first case because $B_{n,1}$ is linear, and in the second case because
$B_{n,2}$ is a commutator and $\rho$ is tracial.  It follows that
$$\lim_{N \to \infty} \rho(\tC_n) = 0.$$

\begin{lemma}
$$\lim_{N \to \infty} \cL[\tvC] = \cL[\vX],$$
where $\vX$ is Gaussian with covariance matrix
$$M_{j,k} = E[X_j X_k] = \rho(B_{j,1} B_{k,1}).$$
\end{lemma}

In other words,
$$\cL[\tC_1,\tC_2,\ldots,\tC_b]$$
approaches the same classical limit as
$$\cL[\tB_{1,1},\tB_{2,1},\ldots,\tB_{b,1}].$$

\begin{proof} Observe that
$$\lim_{N \to \infty} C_n = B_{n,1}.$$

The lemma follows from the proof of \thm{th:ojcentral}, where it was left
unstated that all estimates are locally uniform in $\vA$. We substitute $\vC
- \rho(\vC)$ for $\vA$ in the theorem and use the fact that $\rho(\tC_n) \to
0$.
\end{proof}

The rest of the proof of \thm{th:main} follows the same argument as the case
when $\vB$ is linear by substituting $\vC$ for $\vB$.

\section{Not the proof}

To understand the proof of \thm{th:main}, it may help to see why some
alternative lines of argument do not suffice.

The Giri-von-Waldenfels central limit theorem implies that the moments of
$p(\tvA)$ converge to the moments of $p(\vX)$ by multilinear expansion.
However, unless $p$ is either linear or positive-definite quadratic, $p(\vX)$
is not uniquely determined by its moments because the tail of $\cL[p(\vX)]$ is
too thick.  However, it is yet possible that \thm{th:main} would follow from
the Giri-von-Waldenfels theorem together with an analytic theory of
non-commutative moments.

The Goderis-Verbeure-Vets central limit theorem establishes the convergence
of any
$$\rho^{\tensor N}(e^{i\vt \cdot \tvA}),$$
which can be interpreted as a joint characteristic of $\tvA$. Indeed
\lem{th:ojcentral} more generally establishes the convergence of
$$\rho^{\tensor N}(e^{i\vt_1 \cdot \tvA}e^{i\vt_2 \cdot \tvA}\cdots
    e^{i\vt_b \cdot \tvA})$$
for any sequence of vectors $\vt_1,\vt_2,\ldots,\vt_b$. However, absent an
analytic theory of non-commutative characteristic functions, this does not
imply \thm{th:main}.

The Baker-Campbell-Hausdorff expansion also shows that
$$e^{it_1 \tB_1^{\beta_1}} e^{it_2 \tB_2^{\beta_2}} \cdots
    e^{it_b \tB_b^{\beta_b}}$$
approximates
$$e^{i(t_1 \tB_1^{\beta_1} + t_2 \tB_2^{\beta_2} + \cdots +
    t_b \tB_b^{\beta_b})}.$$
Unfortunately the infinite sum of the BCH expansion does not (as far we know)
commute with the limit $N \to \infty$.

If $\rho$ is not tracial, the proof of \lem{l:spec} fails for the non-spectral
norm
$$||A||_\rho = \rho(A^*A).$$
Alternatively, if $\sigma$ is a tracial state, then
$$||\rho^{\tensor N}||_{\sigma^{\tensor N}} = ||\rho||_\sigma^N \to \infty$$
exponentially, assuming that $\cA$ even has a tracial state and that
$||\rho||_\sigma$ is finite. \lem{l:cint} generalizes to higher-order
commutators which decay more and more quickly, but these suffer from the same
exchange-of-limits problem as the BCH expansion when combined with the norm of
$\rho^{\tensor N}$.  \thm{th:ojcentral} also depends on the assumption
that $\rho$ is tracial, but there the assumption may be unnecessary.

Finally when $\rho$ is not tracial, then $||\cdot||_\rho$ is at least
left-invariant under unitary multiplication. If each commutator $[A_j,A_k]$ is
central, then we can move the unitary factors in \lem{l:cint} to the left and
consequently establish \lem{l:spec} for $||\cdot||_\rho$.  This proves a slight
generalization of \thm{th:main}.  However, it is equivalent to the
Cushen-Hudson-Quaegebeur central limit theorem
\cite{CH:mechanical,Quaegebeur:ccr}, except that they do not require bounded
random variables.


\providecommand{\bysame}{\leavevmode\hbox to3em{\hrulefill}\thinspace}

\end{document}